\documentclass[a4paper,11pt]{article}
\usepackage{pos}

\title{Processing Images from Multiple IACTs in the TAIGA Experiment with Convolutional Neural Networks}
\ShortTitle{Processing Images from Multiple TAIGA IACTs with CNNs}

\author*{Stanislav Polyakov}
\author{Andrey Demichev}
\author{Alexander Kryukov}
\author{Evgeny Postnikov}

\affiliation{Skobeltsyn Institute of Nuclear Physics, M.V. Lomonosov Moscow State University, \\
  1(2) Leninskie gory, Moscow 119991, Russian Federation}

\emailAdd{s.p.polyakov@gmail.com}
\emailAdd{demichev@theory.sinp.msu.ru}
\emailAdd{kryukov@theory.sinp.msu.ru}
\emailAdd{evgeny.post@gmail.com}

\abstract{Extensive air showers created by high-energy particles interacting with the Earth atmosphere can be detected using imaging atmospheric Cherenkov telescopes (IACTs). The IACT images can be analyzed to distinguish between the events caused by gamma rays and by hadrons and to infer the parameters of the event such as the energy of the primary particle. We use convolutional neural networks (CNNs) to analyze Monte Carlo-simulated images from the telescopes of the TAIGA experiment. The analysis includes selection of the images corresponding to the showers caused by gamma rays and estimating the energy of the gamma rays. We compare performance of the CNNs using images from a single telescope and the CNNs using images from two telescopes as inputs.}

\FullConference{%
  The 5th International Workshop on Deep Learning in Computational Physics\\
  28-29 June, 2021\\
  Moscow, Russia
}

\begin{document}
\maketitle

\section{Introduction}

An extensive air shower caused by a high-energy particle (cosmic or gamma ray) interacting with upper atmosphere can be detected by several methods including imaging atmospheric Cherenkov telescopes (IACTs). In Russian TAIGA (Tunka Advanced Instrument for cosmic ray physics and Gamma-ray Astronomy) experiment the number of installed and commissioned IACTs has been increased from one to two in 2020, and the third telescope was installed in 2020 \cite{Postnikov}. 

Convolutional neural networks (CNNs) are a very successful machine learning tool. Several research teams have demonstrated high performance of CNNs for the analysis of images from IACTs and IACT arrays of several gamma astronomy experiments such as VERITAS \cite{VERITAS}, CTA \cite{CTA}, H.E.S.S. \cite{HESS}. We previously applied CNNs to the analysis of images from a single TAIGA IACT, specifically, to the problems of identification of the event types and estimation of the energy of the original gamma rays \cite{APPDS1, APPDS2}.

In this paper we apply convolutional neural networks to the identification of the event types and estimation of the energy of the original gamma rays based on images from one or two TAIGA Cherenkov telescopes and compare the neural network performance in monoscopic and stereoscopic modes.

\section{Particle identification}

We prepared a dataset S1 of 3400 gamma events and 9306 proton events recorded by two IACTs positioned at a varying distance between 300 m and 350 m from each other. The dataset was generated using Monte Carlo simulation software CORSIKA \cite{CORSIKA}.

\begin{figure}
\centering
\includegraphics[width=1.0\linewidth]{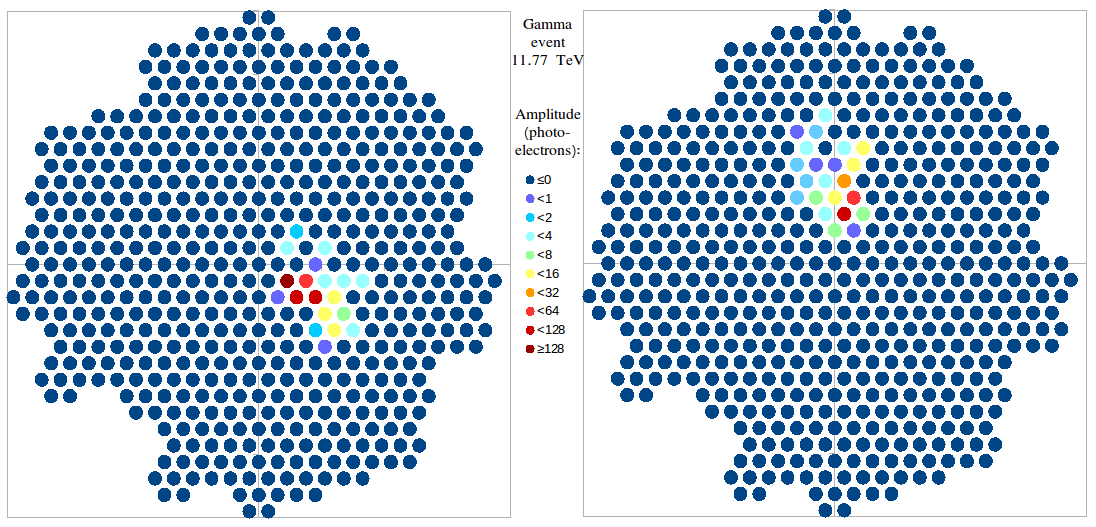}
\caption{Simulated stereoscopic TAIGA IACT image of a gamma event, E = 11.77 TeV.}
\label{fig:gamma}
\end{figure}

\begin{figure}
\centering
\includegraphics[width=1.0\linewidth]{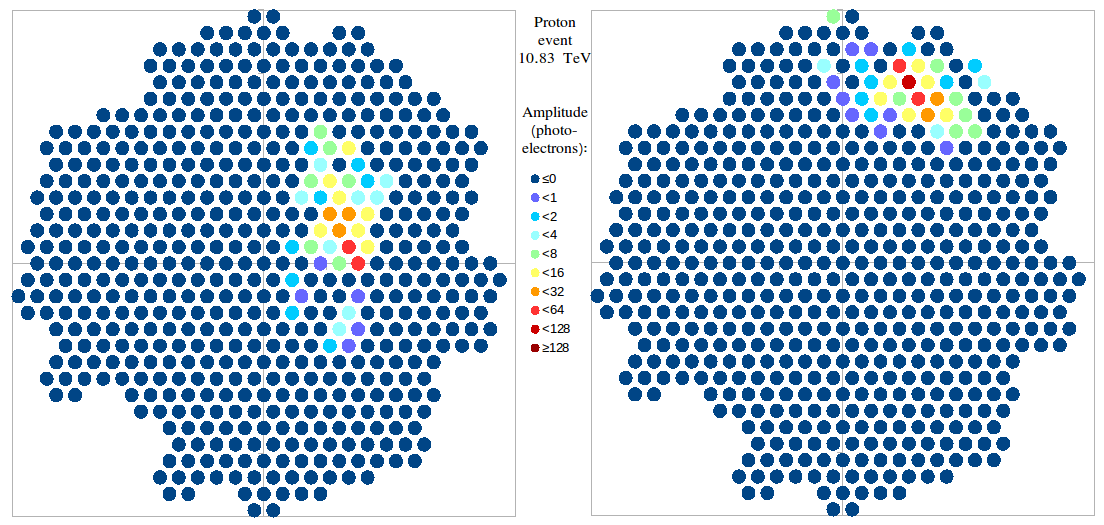}
\caption{Simulated stereoscopic TAIGA IACT image of a proton event, E = 10.83 TeV.}
\label{fig:proton}
\end{figure}

We used the following neural network architecture with $W = 10, 15, 25, 50, 100$:
\begin{center}
Conv2D $5\times5$, $W$ \\*
AvgPool $2\times2$ \\*
Conv2D $5\times5$, $W$ \\*
AvgPool $2\times2$ \\*
Conv2D $3\times3$, $W$ \\*
AvgPool $2\times2$ \\*
Flatten $3\times3\times W  \rightarrow 9 W$ \\*
Fully connected layer, $3W$ \\*
Fully connected layer, $W$ \\*
Output layer, $2$
\end{center}

The neural networks were implemented using PyTorch library %\cite{PyTorch} 
and trained using the events from the dataset S1 in 2 modes: monoscopic (only using the image from the first telescope) and stereoscopic (using images from both telescopes). On each iteration, 80\% of the dataset were used for training, and the remaining 20\% for the evaluation of results. We then calculated the average results between 10 iterations.

Dropout with probability 50\% was applied during training before each fully connected layer, including the output layer. The neural networks were trained using gradient descent for 500 epochs. The initial learning rate was 0.1 and whenever the average loss for the training set did not decrease below the last minimum for over 20 epochs, the learning rate was decreased by a factor of 10.

\begin{figure}
\centering
\includegraphics[width=1.0\linewidth]{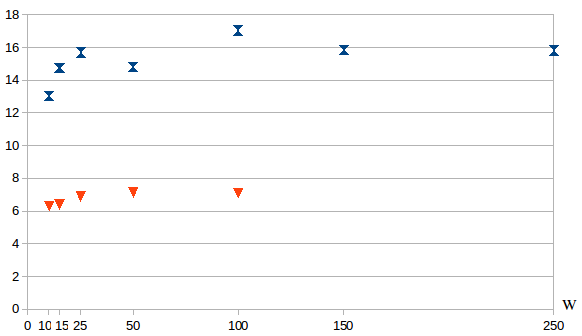}
\caption{$Q$ factor for CNNs using one (triangles) and two (double triangles) IACT images as input.}
\label{fig:classification}
\end{figure}

As the measure of quality of particle identification we calculated the selection quality factor $Q$, which indicates an improvement of a significance of the statistical hypothesis that the events do not belong to the background in comparison with the significance before selection. For Poisson distribution
$$
Q = \frac{\Gamma_{true}/\Gamma}{\sqrt{\Gamma_{false}/H}},
$$
where $\Gamma$ and $H$ are the total number of gamma events and background hadron events, respectively, and $\Gamma_{true}$ and $\Gamma_{false}$ are the number of events correctly and incorrectly identified as gamma events. The results are shown on the Fig. \ref{fig:classification}. 

An event from a test set was identified as a gamma event if the CNN estimated its likelihood of being a gamma event above the threshold sufficient to retain 60\% of the gamma events in the training set. The highest value of $Q$ factor for monoscopic CNNs we found was 7.1, and the highest value for stereoscopic CNNs was 17.0. The additional data from the second telescope results in the decrease of the number of misidentified proton events by a factor of 4--5 without decreasing the number of correctly identified gamma events. 

\section{Gamma ray energy estimation}

For training and evaluating convolutional neural networks estimating the energy of gamma rays we primarily used a Monte Carlo-generated dataset S2 of 18359 gamma events detected by two IACTs with a distance 324 m between them. The energy of the gamma rays was between 1 and 50 TeV. Each convolutional network was trained 10 times on a randomly chosen 80\% subset of the dataset. 

The neural networks were trained in monoscopic and telescopic mode using 80\% of the dataset for training and the remaining 20\% for the evaluation of results. We then calculated the average results between 10 iterations.

Dropout with probability $p$ was applied during training before each fully connected layer, including the output layer. For each neural network we found the value of $p = 0.025k$, $k = 1, 2, \ldots, 20$, that resulted in the most accurate energy estimate. 

The neural networks were trained using gradient descent. The initial learning rate was 0.1 and when the average loss for the training set did not decrease below the last minimum for over 20 epochs, the learning rate was decreased by a factor of 10. If the learning rate decreased below 0.001, the training stopped. The hard limit was set to 2000 epochs, but it was never reached.

An example of neural network architecture for energy estimation ($W = 25, 50, 100$ were used):
\begin{center}
Conv2D $5\times5$, $\lfloor W/4 \rfloor$ \\*
AvgPool $2\times2$ \\*
Conv2D $5\times5$, $\lfloor W/2 \rfloor$ \\*
AvgPool $2\times2$ \\*
Conv2D $3\times3$, $\lfloor W/2 \rfloor$ \\*
AvgPool $2\times2$ \\*
Flatten $3\times3\times \lfloor W/2 \rfloor  \rightarrow 9 \lfloor W/2 \rfloor $ \\* 
Fully connected layer, $W$ \\*
Fully connected layer, $W$ \\*
Fully connected layer, $W$ \\*
Output layer, 1
\end{center}

The most accurate monoscopic neural networks that we found estimated the energy of gamma rays in S2 dataset with the average relative error 24.0\%. The most accurate stereoscopic neural networks estimated the energy of S2 gamma events with the average relative error 12.5\%.

\begin{figure}
\centering
\includegraphics[width=1.0\linewidth]{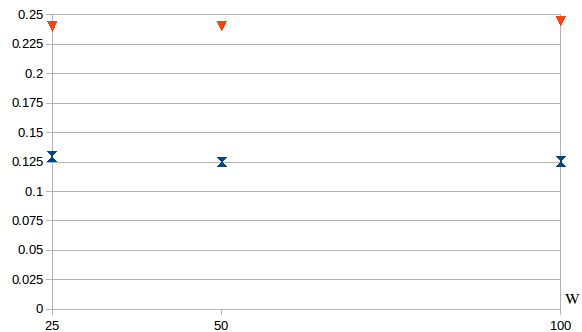}
\caption{Average relative error of energy estimates in monoscopic (triangles) and stereoscopic (double triangles) modes.}
\label{fig:estimation}
\end{figure}

We also trained several neural networks on the gamma events from the dataset S1 (energy between 1 and 45 TeV). The best results for average relative error were 20.8\% in monoscopic mode and 15.5\% in stereoscopic mode.

\section{Conclusions}

We applied convolutional neural networks to the problems of particle type identification and gamma ray energy estimation based on IACT images in monoscopic and stereoscopic mode. We found that adding an image from a second telescope as input to the same neural network can increase the cosmic ray background suppression by a factor of 4--5, and decrease the average relative error of the energy estimates by a factor of 1.3--2.

\acknowledgments{This work was carried out in the framework of R\&D State Assignment No.115041410196.}

\bibliographystyle{JHEP}
\bibliography{bib-polyakov}

\providecommand{\href}[2]{#2}\begingroup\raggedright\begin{thebibliography}{1}

\bibitem{Postnikov}
E.B.~Postnikov, I.I.~Astapov, P.A.~Bezyazeekov, M.~Blank and A.N.~Borodin,
  \emph{First detection of gamma-ray sources at tev energies with the first
  imaging air cherenkov telescope of the taiga installation},
  \href{https://doi.org/10.1088/1742-6596/1690/1/012023}{\emph{J. Phys. Conf.
  Ser.} {\bfseries 1690} (2020) 012023}.

\bibitem{VERITAS}
Q.~Feng, J.~Jarvis, V.~Collaboration et~al., \emph{A citizen-science approach
  to muon events in imaging atmospheric cherenkov telescope data: the muon
  hunter},  in \emph{35th International Cosmic Ray Conference (ICRC2017)},
  vol.~301, 2017.

\bibitem{CTA}
S.~Mangano, C.~Delgado, M.I.~Bernardos, M.~Lallena, J.J.R.~V{\'a}zquez,
  C.~Consortium et~al., \emph{Extracting gamma-ray information from images with
  convolutional neural network methods on simulated cherenkov telescope array
  data},  in \emph{IAPR Workshop on Artificial Neural Networks in Pattern
  Recognition}, pp.~243--254, Springer, 2018.

\bibitem{HESS}
I.~Shilon, M.~Kraus, M.~B{\"u}chele, K.~Egberts, T.~Fischer, T.~Holch et~al.,
  \emph{Application of deep learning methods to analysis of imaging atmospheric
  cherenkov telescopes data},
  \href{https://doi.org/10.1016/j.astropartphys.2018.10.003}{\emph{Astroparticle
  Physics} {\bfseries 105} (2019) 44}.

\bibitem{APPDS1}
E.~Postnikov, A.~Kryukov, S.~Polyakov, D.~Shipilov and D.~Zhurov,
  \emph{Gamma/hadron separation in imaging air cherenkov telescopes using deep
  learning libraries tensorflow and pytorch},
  \href{https://doi.org/10.1088/1742-6596/1181/1/012048}{\emph{J. Phys. Conf.
  Ser.} {\bfseries 1181} (2019) 012048}.

\bibitem{APPDS2}
E.~Postnikov, A.~Kryukov, S.~Polyakov and D.~Zhurov, \emph{Deep learning for
  energy estimation and particle identification in gamma-ray astronomy},
  vol.~2406, pp.~90--99, 2019.

\bibitem{CORSIKA}
D.~Heck, J.~Knapp, J.N.~Capdevielle, G.~Schatz and T.~Thouw, \emph{Corsika: A
  monte carlo code to simulate extensive air showers},  Tech. Rep. FZKA-6019,
  Karlsruhe (2, 1998).

\end{thebibliography}\endgroup

\end{document}